\documentclass[aps,prd,showpacs,preprintnumbers,twocolumn,groupedaddress,nofootinbib]{revtex4}
\usepackage{graphicx}
\usepackage{latexsym}
\def\beq{\begin{equation}}
\def\eeq{\end{equation}}
\def\bey{\begin{eqnarray}}
\def\eey{\end{eqnarray}}

\def\sigv{\langle\sigma v\rangle}

\def\lsim{\mathrel{\raise.3ex\hbox{$<$\kern-.75em\lower1ex\hbox{$\sim$}}}}
\def\gsim{\mathrel{\raise.3ex\hbox{$>$\kern-.75em\lower1ex\hbox{$\sim$}}}}

\begin{document}

\title{Gamma Rays From The Galactic Center and the WMAP Haze}  
\author{Dan Hooper$^{1,2}$ and Tim Linden$^{1,3}$}
\affiliation{$^1$Center for Particle Astrophysics, Fermi National Accelerator Laboratory, Batavia, IL 60510}
\affiliation{$^2$Department of Astronomy and Astrophysics, University of Chicago, Chicago, IL 60637}
\affiliation{$^3$Department of Physics, University of California, Santa Cruz, CA 95064}

\date{\today}

\begin{abstract}
Recently, an analysis of data from the Fermi Gamma Ray Space Telescope has revealed a flux of gamma rays concentrated around the inner $\sim$$0.5^{\circ}$ of the Milky Way, with a spectrum that is sharply peaked at 2-4 GeV. If interpreted as the products of annihilating dark matter, this signal implies that the dark matter consists of particles with a mass between 7.3 and 9.2 GeV annihilating primarily to charged leptons. This mass range is very similar to that required to accommodate the signals reported by CoGeNT and DAMA/LIBRA. In addition to gamma rays, the dark matter is predicted to produce energetic electrons and positrons in the Inner Galaxy, which emit synchrotron photons as a result of their interaction with the galactic magnetic field. In this letter, we calculate the flux and spectrum of this synchrotron emission assuming that the gamma rays from the Galactic Center originate from dark matter, and compare the results to measurements from the WMAP satellite. We find that a sizable flux of hard synchrotron emission is predicted in this scenario, and that this can easily account for the observed intensity, spectrum, and morphology of the ``WMAP Haze''.
\end{abstract}

\pacs{95.35.+d;95.30.Cq,95.55.Ka; FERMILAB-PUB-10-472-A}
\maketitle


Dark matter in the form of weakly interacting massive particles (WIMPs) is predicted in many models to produce observable signals such as gamma-rays, neutrinos, charged particles and synchrotron radiation through its annihilations~\cite{syn}. Recently, an analysis of the first two years of data from the Fermi Gamma Ray Space Telescope has identified a component of gamma-ray emission that is highly concentrated within 0.5$^{\circ}$ around the galactic center, and which is spectrally and morphologically distinct from all known backgrounds~\cite{newlisa}. While a new astrophysical source such as an undetected population of millisecond pulsars may create such a signal~\cite{gcmsps}, the morphology of this emission can most easily be accounted for by dark matter with a mass of 7.3-9.2 GeV annihilating primarily to tau leptons (possibly in addition to other leptons), and distributed in a cusped, and possibly adiabatically contracted \cite[see e.g.][]{gnedin_primack}, profile (with $\rho \propto r^{-1.18}$ to $\rho \propto r^{-1.33}$). The normalization of this observed signal requires the dark matter annihilation cross section (to $\tau^+ \tau^-$) to fall within the range of  $\sigv = 4.3 \times 10^{-27}$ to $3.9 \times 10^{-26}$ cm$^3$/s, similar to that required of a thermal relic~\cite{newlisa,models}. The mass range indicated by this observation is similar to that required to accommodate the excess events reported by the CoGeNT~\cite{cogent} collaboration, and the annual modulation observed by DAMA/LIBRA~\cite{dama,consistent}. Recently, Buckley et al.~\cite{models} have shown that the phenomenology of this leptophilic dark matter model is consistent with known constraints from the relic abundance, collider physics and direct dark matter detection.

If leptophilic dark matter is responsible for the gamma ray signal observed in the galactic center, then a sizable flux of energetic electrons/positrons will necessarily be injected into the inner galaxy. Such particles lose energy via synchrotron, inverse Compton and bremsstrahlung. Given that the dark matter's mass, profile, annihilation cross section, and annihilation channels are highly constrained by the intensity, morphology, and spectral shape of the observed gamma rays from the galactic center, it is possible to make a robust prediction for the characteristics of the corresponding synchrotron signal, providing us the opportunity to either falsify or further support a dark matter interpretation of the gamma ray signal. 

GeV-scale cosmic ray electrons/positrons propagating in 10-100$\mu$G magnetic fields produce synchrotron radiation that peaks at GHz frequencies. The emission is thus naturally expected to fall within the frequency range studied by cosmic microwave background (CMB) missions, such as the Wilkinson Microwave Anisotropy Probe (WMAP)~\cite{spergel}. Data from WMAP and other CMB experiments can, therefore, be used to potentially constrain or detect the occurrence of dark matter annihilations in our galaxy. Interestingly, WMAP observations have revealed an excess of microwave emission in the inner $20^{\circ}$ around the center of the Milky Way, distributed with approximate radial symmetry~\cite{haze1}. Although this excess is controversial~\cite{gold2011}, and may have an intensity which is affected by the details of astrophysical foreground subtraction~\cite{mertsch}, the spectrum and morphology of this source does not correlate with any known foreground~\cite{doblerfink}. Possible astrophysical origins, such as thermal bremsstrahlung (free-free emission) from hot gas, thermal dust, spinning dust, and Galactic synchrotron traced by low-frequency surveys have proven problematic~\cite{haze1,doblerfink}. Alternatively, it has been suggested that the haze could be synchrotron emission from a distinct population of electrons and positrons produced by dark matter particles annihilating in the inner galaxy~\cite{darkhaze,darkhaze1}.

The possibility that light ($\sim$5-10 GeV) annihilating dark matter particles may explain the presence of the WMAP haze has been bolstered by recent claims that the magnetic field near the Galactic Center may be significantly stronger than previously thought. In particular, Crocker et al.~\cite{nature} analyzed the spectral break in the Inner Galaxy's non-thermal radio spectrum and set a lower limit of 50$\mu$G on the magnetic field within 400~pc of the Galactic Center. Further work~\cite{crocker2} found a best fit for 100-300$\mu$G magnetic fields in the region around the Galactic Center. Such large magnetic fields in the Inner Galaxy would result in a brighter flux of synchrotron emission and a harder synchrotron spectral index than had previously been expected. 




In this study, we will address the question of whether the observed synchrotron emission from the inner kiloparsecs of the Milky Way is consistent with the a dark matter interpretation of the gamma ray signal from the Galactic Center. The galactic dark matter distribution is thought to be consistent with an NFW profile~\cite{nfw}, and in the case of the WMAP haze, it has been shown that small deviations from this distribution (such as an Einasto or Via Lactea profile), have little effect on the morphology of the WMAP haze~\cite{lindenwmap}. However, in order to match the morphology of the $\gamma$-ray signal near the galactic center, we require a dark matter density profile which is steeper than a traditional r$^{-1}$ employed in~\cite{nfw}. We note that such a contraction (to as steep as r$^{-1.5}$) is predicted by models of adiabatic contraction due to baryonic interactions in the inner galaxy~\cite{gnedin_primack}. We note, however, that the position of the spectral break between the adiabatic contracted inner galaxy and the r$^{-1}$ falloff of nearer the solar position is uncertain, and thus we consider two different dark matter halo profiles to bracket this uncertainty. First, we adopt a simple extrapolation of the distribution implied by the $\gamma$-rays from the galactic center (halo profile A):

\begin{equation}
\rho_A(r) = D_0 \,{\rm GeV/cm^3} \, \bigg(\frac{8.5 \,{\rm kpc}}{r}\bigg)^{\alpha},
\end{equation}
where $r$ is the distance from the Galactic Center, and we evaluate $\alpha$ = \{1.18, 1.34\}. The normalization value D$_0$ has been set such that the total mass within the solar circle is equal to that obtained with a $r^{-1}$ profile and a local density of 0.3 GeV/cm$^3$. For the value $\alpha$~=~1.18 we have D$_0$~=~0.273~GeV/cm$^3$, and for $\alpha$~=~1.33 we obtain D$_0$~=~0.251~GeV/cm$^3$. Additionally, we consider a broken power-law profile (halo profile B):
\begin{eqnarray}
\rho_B(r) &=& 0.30 \, {\rm GeV/cm^3} \bigg(\frac{8.5 \, {\rm kpc}}{r}\bigg)^{1.0}, \,\,\, r>0.5\, {\rm kpc} \,\,\,\,\,\,\, \nonumber \\
\rho_B(r) &=& 5.1 \, {\rm GeV/cm^3} \, \bigg(\frac{0.5 \, {\rm kpc}}{r}\bigg)^{1.33}, \,\,\, r<0.5\, {\rm kpc}. \,\,\,\,\,\,\,
\end{eqnarray}
We do not comment here on the case of the $\alpha$~=~1.18 profile with a broken power law profile, as the results are not greatly altered compared to the unbroken profile. To generate the observed gamma ray flux from the Galactic Center, profile A requires an annihilation cross-section of 4.3~x~10$^{-27}$~cm$^3$s$^{-1}$ for the case $\alpha$=1.33 and 3.2~x~10$^{-26}$~cm$^3$s$^{-1}$ for the case $\alpha$=1.18. For profile B, we evaluate only the case $\alpha$=1.33 and set an annihilation cross-section of 1.3~x~10$^{-26}$~cm$^3$s$^{-1}$~\cite{newlisa}.

It is possible that in addition to $\tau^+ \tau^-$, the dark matter may also annihilate to final states which produce few gamma rays, but contribute significantly to the cosmic ray electron/positron spectrum, and therefore to the resulting synchrotron emission. We will consider two cases: one in which the dark matter annihilates uniquely to $\tau^+ \tau^-$, and another in which it annihilates equally to  $e^+ e^-$, $\mu^+ \mu^-$, and $\tau^+ \tau^-$ (which we will refer to as ``democratic leptons''). The latter case injects roughly an order of magnitude more energy into cosmic ray electrons, and with a somewhat harder spectrum than that from taus alone.

To model the propagation of electron and positrons produced through dark matter annihilation, we employ the cosmic ray propagation code $Galprop$~\cite{galprop1, galprop2}, which calculates the synchrotron energy spectrum including effects such as diffusion, reacceleration, and alternative energy loss mechanisms including inverse Compton scattering and bremsstrahlung radiation. We have modified the $Galprop$ program to accept a dark matter annihilation spectrum from the DarkSUSY package~\cite{darksusy}, which has been formulated to produce the lepton fluxes for a variety of WIMP annihilation channels. Unless otherwise noted, we have adopted parameters identical to the best fit propagation model (base) determined in Ref.~\cite{lindenwmap}, which were found to provide the best agreement to observed cosmic ray primary to secondary ratios. We adopt a diffusion constant D$_0$~=~5.0~x~10$^{28}$~cm$^2$s$^{-1}$, an Alfv\`en velocity v$_\alpha$~=~25~km~s$^{-1}$, and assume negligible convection.

We parametrize the Galactic Magnetic Field according to:
\begin{equation}
B(r,z) = B_0 \, e^{-(r-1\,{\rm kpc})/R_s} \, e^{-(|z|-2\,{\rm kpc})/z_s},
\end{equation}
where $r$ and $z$ represent the distance from the Galactic Center along and perpendicular to the the Galactic Plane. This exponential distribution follows the standard set forth in~\citep{galprop1, galprop2}, and we employ values of R$_s$ and z$_s$ in order to approximately match the magnetic field distribution both near the galactic center set by~\citep{nature}, and locally set by~\citep{localB}. We note that these parameters suffer from significant uncertainties, and we demonstrate the impact of these uncertainties on our models below.

In order to extract the WMAP haze residual from the WMAP dataset, we utilize the foreground subtraction templates employed in Ref.~\citep{doblerfink} using the template subtraction method CMB5, and we complete the mean subtraction of the background dataset as in Ref.~\citep{lindenwmap}. Since the dominant source of error in the determination of the WMAP haze stems from the accuracy of these subtraction methods - rather than from the measured error in the WMAP dataset, we resist calculating error bars for our measurement, as they would serve primarily to belie to actual error in the WMAP haze measurement (for a quantifications of these errors, see Fig. 8 in \cite{doblerfink}). Instead, we include one-~$\sigma$ error bars based on the temperature fluctuations in each radial and energy bin. In order to model the isotropic component which was subtracted from the WMAP dataset in the template fits, we add an isotropic component to the simulated haze which is allowed to float individually in each energy band (labeled the zero point offset). This is intended to account for systematic uncertainties in the template subtractions, as well as for any other astrophysical sources of hard synchrotron that might be present.

\begin{figure}\mbox{\includegraphics[width=0.5\textwidth,clip]{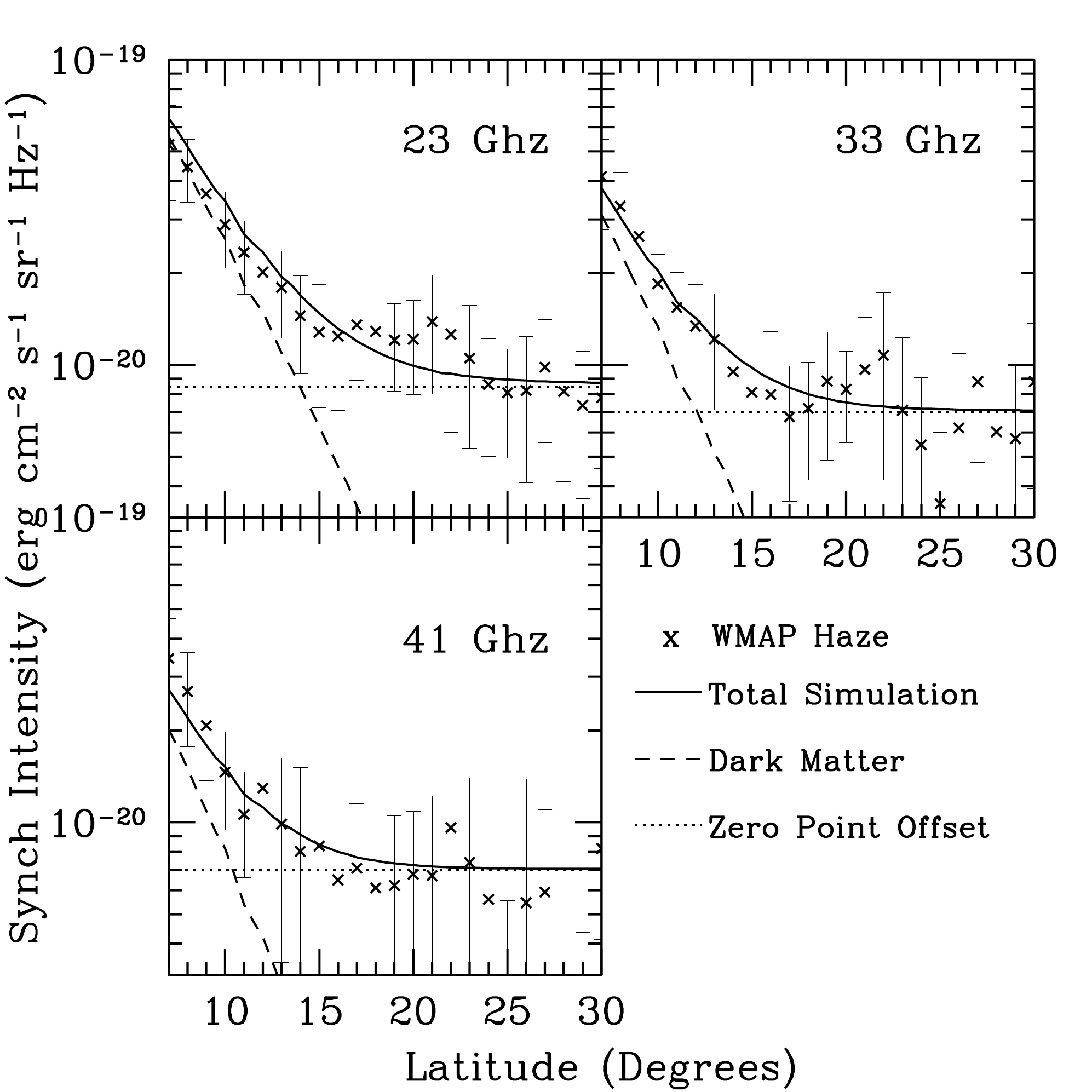}}\caption{Synchrotron emission from dark matter annihilations as a function of latitude below the Galactic Center for 8 GeV dark matter particles annihilating to $\tau^+ \tau^-$, and using halo profile A with a slope $\alpha$~=~1.18. The total cross section of $\sigma$v~=~3.2~x~10$^{-26}$ cm$^3$/s was chosen to normalize the gamma ray signal observed from the Galactic Center. The magnetic field model used is given by B(r,z)~=~27.7\,$\mu$G \,\,$e^{-(r-1\,{\rm kpc})/3.2 \,{\rm kpc}}$~\,$e^{-(|z|-2\,{\rm kpc})/1.0\, {\rm kpc}}$. The ``dark matter" signal corresponds to the synchrotron signal provided by dark matter leptons in our simulations, while the total signal includes the ``dark matter" signal added to the zero-point offset. The error bars shown correspond to the 1-$\sigma$ temperature fluctuations in each radial and energy bin~\cite{doblerfink}}
\label{tau_a_118}
\end{figure}

\begin{figure}\mbox{\includegraphics[width=0.5\textwidth,clip]{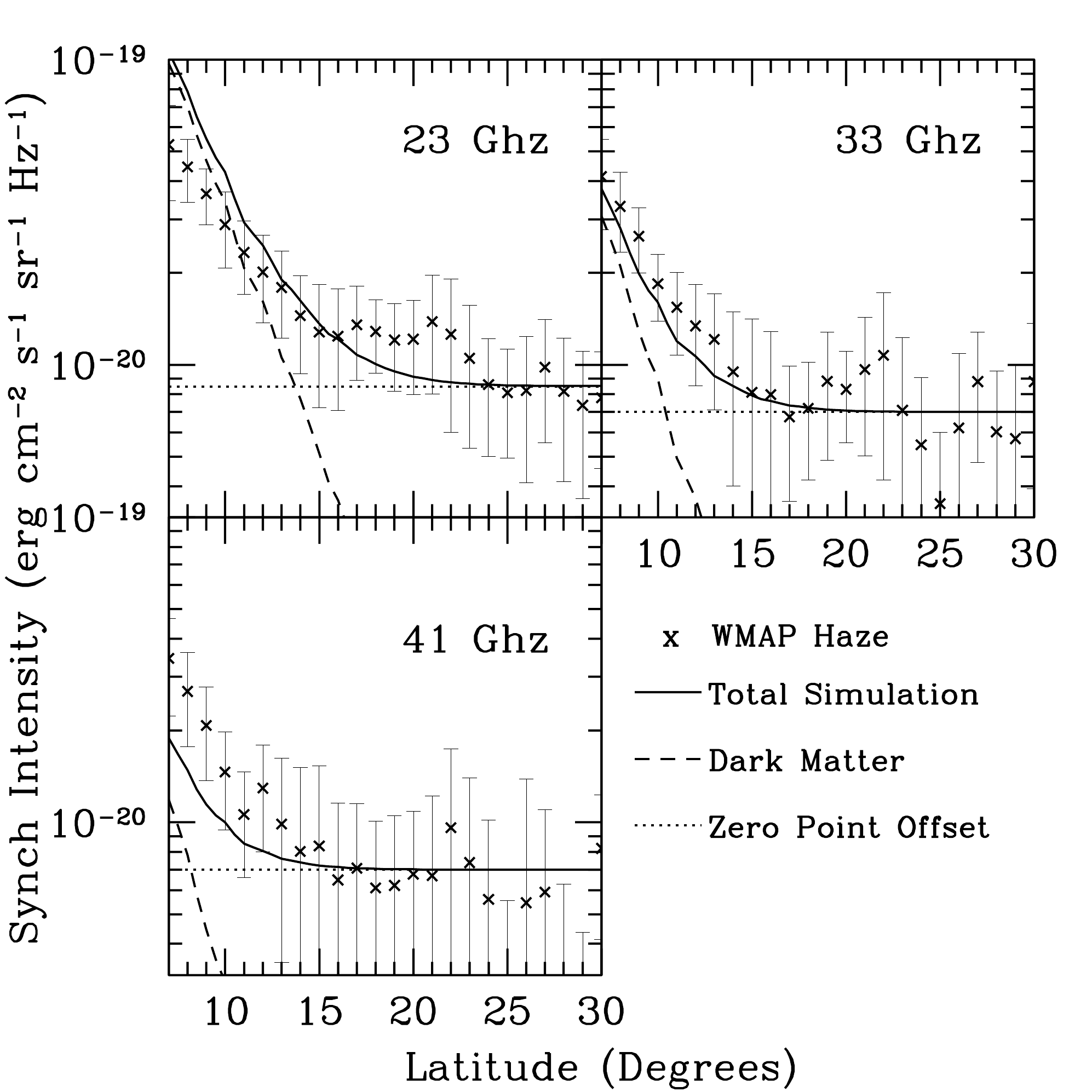}}\caption{As in Fig.~\ref{tau_a_118}, for a dark matter model which annihilates democratically into leptons with a cross section of $\sigma$v~=~9.6~x~10$^{-26}$ cm$^3$s$^{-1}$ (again, normalized to the observed gamma ray signal). The magnetic field model used is given by B(r,z)~=~6.28\,$\mu$G \,\, $e^{-(r-1\,{\rm kpc})/4.0\,{\rm kpc}}$~\,$e^{-(|z|-2\,{\rm kpc})/2.2\,{\rm kpc}}$.}
\label{democratic_a_118}
\end{figure}

Our models indicate that the details of the annihilation pathway postulated by \cite{newlisa} are the most important factor in controlling the dark matter match to the WMAP haze. We find that the $\tau$-only annihilation channel is not able to reproduce the intensity of the WMAP haze using $\alpha$~=~1.33 with either profile- even if we allow for an unrealistically large magnetic field of 180$\mu$G in the haze region. However, In the case of an $\alpha$~=~1.18, we are assisted by the much larger cross-section of 3.2~x~10$^{-26}$ cm$^3$s$^{-1}$ required to match the observed $\gamma$-ray signal. In Figure~\ref{tau_a_118} we show that this allows for a reasonable match to the WMAP haze if we allow for strong magnetic fields on the order of  28$\mu$G in the haze region. We find that the results are changed only slightly if we introduce a spectral break for this profile. 

In the case where dark matter instead annihilates democratically into leptons, we obtain a strong enhancement to the intensity of the WMAP haze for two reasons: (1) the cross-sections are multiplied by three because we obtain negligible direct $\gamma$-ray production from the muon and electron channels, so it is only the $\tau$ cross section which must be held constant in~\cite{newlisa}, (2) the electron channel converts energy entirely into leptons as opposed to the much smaller contributions from the $\tau$ and $\mu$ channels. In order to account for this much larger input lepton density, we must greatly decrease the magnetic field to maintain a constant synchrotron intensity. In Figure~\ref{democratic_a_118} we show the match for profile A with $\alpha$~=~1.18, which employs a magnetic field of strength 6.3$\mu$G in the Haze region.

\begin{figure}
\mbox{\includegraphics[width=0.5\textwidth,clip]{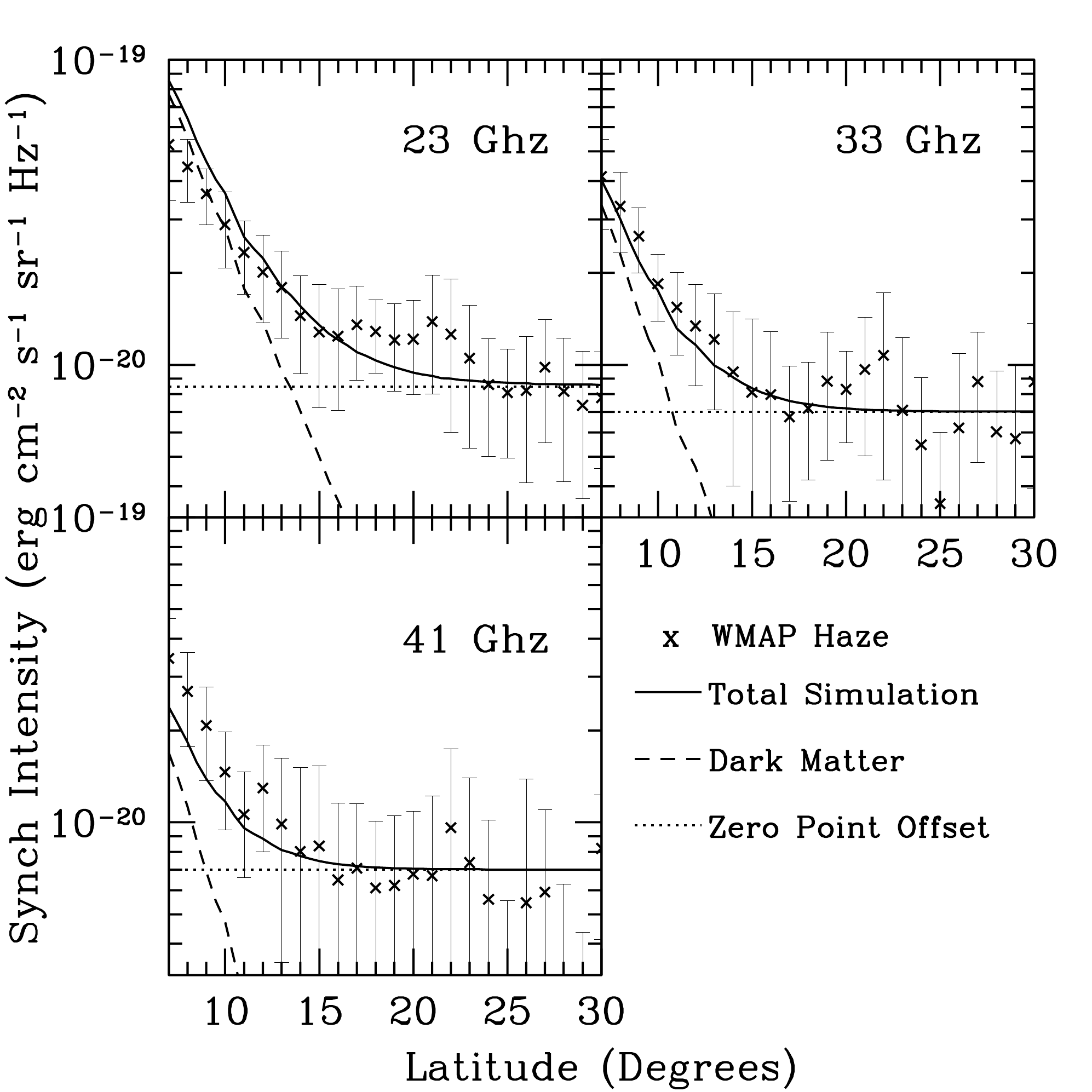}}
\caption{As in Fig.~\ref{democratic_a_118} for a dark matter profile with $\alpha$~=~1.33 and profile A. The total annihilation cross-section has been normalized to $\sigma$v~=~1.3~x~10$^{-26}$ cm$^3$s$^{-1}$ The magnetic field model is given by B(r,z)~=~9.43\,$\mu$G \,\, $e^{-(r-1\,{\rm kpc})/5.0\,{\rm kpc}}$~\,$e^{-(|z|-2\,{\rm kpc})/1.8\,{\rm kpc}}$.}
\label{democratic_a_133}
\end{figure}

\begin{figure}
\mbox{\includegraphics[width=0.5\textwidth,clip]{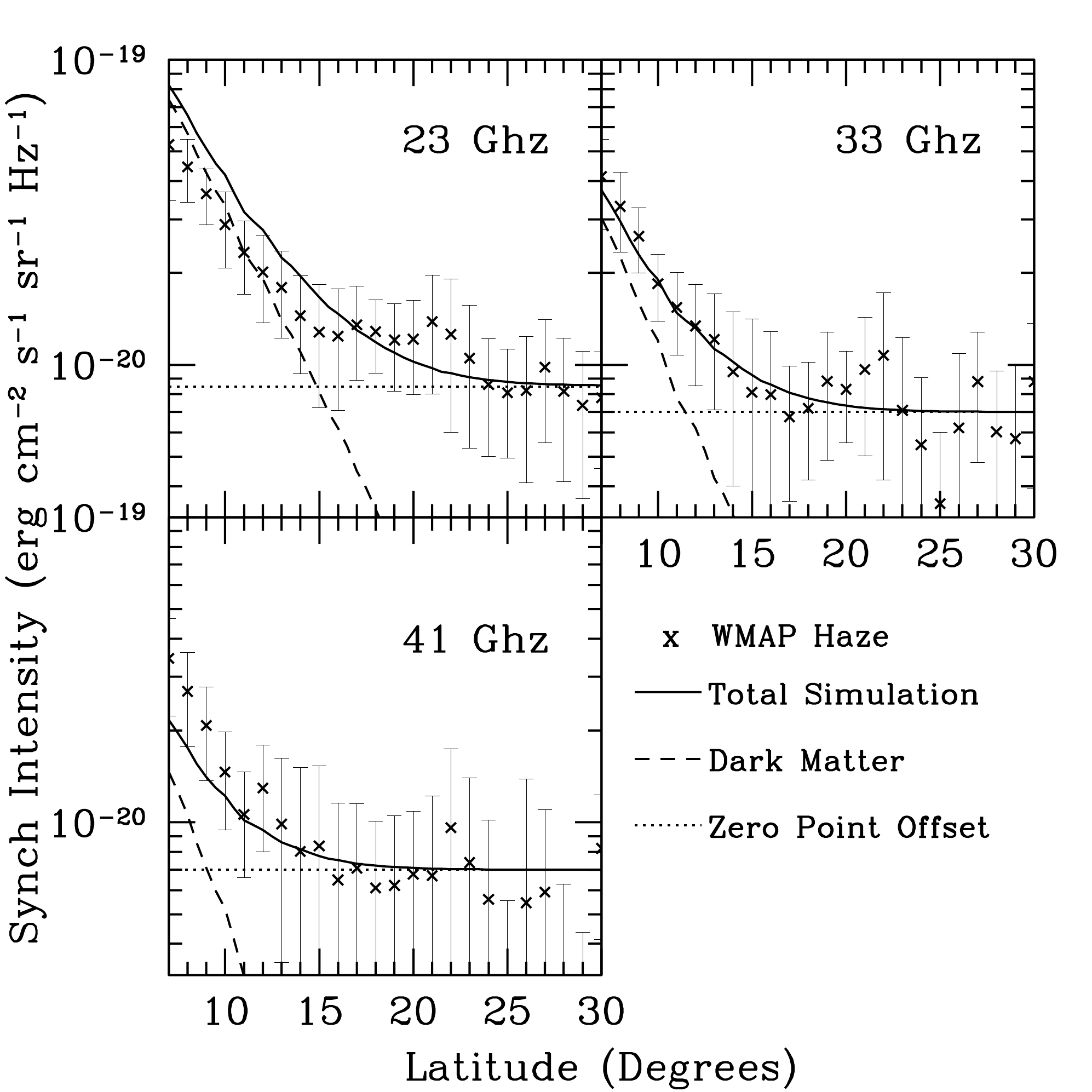}}
\caption{As in Figs.~\ref{democratic_a_133} except with profile B. The total annihilation cross-section has been normalized to $\sigma$v~=~3.9~x~10$^{-26}$ cm$^3$s$^{-1}$ The magnetic field model is given by B(r,z)~=~9.41\,$\mu$G \,\, $e^{-(r-1\,{\rm kpc})/4.0\,{\rm kpc}}$~\,$e^{-(|z|-2\,{\rm kpc})/2.2}$}
\label{democratic_b_133}
\end{figure}

For the case of $\alpha$~=~1.33, we can observe the small differences between profile A and profile B. In Figure~\ref{democratic_a_133} we show the best fit dark matter profile for profile A with a cross-section of 1.3~x~10$^{-26}$ cm$^3$s$^{-1}$, using a magnetic field of strength 9.4$\mu$G in the haze region. This contrasts only slightly with our result in Figure~\ref{democratic_b_133}, where we employ profile B and a cross-section of 3.9~x~10$^{-26}$ cm$^3$s$^{-1}$. Thus we find the effect of both extremes of our dark matter profile extrapolation to have only a minimal effect. 

In summary, a spectrum of gamma rays sharply peaked at 2-4 GeV from the inner $0.5^{\circ}$ around the Galactic Center has recently been identified within the data of the Fermi Gamma Ray Space Telescope~\cite{newlisa}. If these gamma rays are interpreted as dark matter annihilation products, this tightly constrains the dark matter particle's mass, annihilation cross section, annihilation channels, and halo profile. In this letter, we calculate the synchrotron emission from the Inner Galaxy predicted in this scenario, and compare it to the observed features of the WMAP Haze. Although the fits we have shown are not perfect matches to the data, they are well within the expectation given the likely imperfections in simulations of the dark matter profiles, and lepton diffusion employed. We further note that the systematic uncertainties in the determination of the WMAP haze signal exceeds the error bars plotted here, and possibly introduces morphological and spectral uncertainties~\citep{doblerfink}. We find agreement with the observed intensity, spectrum, and morphology of WMAP haze, especially if dark matter is seen to annihilate democratically to leptonic final states (equally to $e^+ e^-$, $\mu^+ \mu^-$, and $\tau^+ \tau^-$). If dark matter annihilations produce only $\tau^+ \tau^-$, we can still match the observed WMAP intensity if we resort to relatively strong magnetic fields in the haze region. We note the magnetic field strengths between these two scenarios differ greatly, and an accurate determination of the magnetic field in this region would constrain the possible dark matter annihilation pathway. 

We emphasize that these fits to the characteristics of the WMAP Haze were obtained with relatively little freedom in the astrophysical or dark matter parameters. In particular, the mass, annihilation cross section, and halo profile are each tightly constrained by the observed features of the Galactic Center gamma ray signal. We note that we have allowed the magnetic field energy density to modulate in order to obtain best fits to the WMAP dataset. Although these magnetic field choices allowed us to adjust the morphology and spectrum of the of the synchrotron emission to a limited degree, we had only moderate ability to adjust the overall synchrotron intensity. Finally, we note that these magnetic field modulations are observationally testable. For this reason, we find it particularly interesting that the dark matter model implied by the observed gamma ray signal so naturally yields synchrotron emission consistent with the WMAP Haze. While these results our not unique to our dark matter models as compared to those of more massive particles (see e.g.~\cite{darkhaze, caceres}), these results show the promise for radio observations to constrain the scenarios of high energy dark matter annihilation such as those shown in \cite{newlisa}, and the need for further studies, such as a model for synchrotron emission near the galactic center where the total $\gamma$-ray intensity is much better constrained.

\smallskip

We would like to thank Greg Dobler for valuable discussions. This work has been supported by the US Department of Energy and by NASA grant NAG5-10842. TL is supported by a Fermilab Fellowship in Theoretical Physics.

\end{document}